\documentclass[pdflatex,sn-mathphys]{sn-jnl}% Math and Physical Sciences Reference Style
\jyear{2021}%

\theoremstyle{thmstyleone}%
\theoremstyle{thmstyletwo}%

\usepackage[displaymath, mathlines]{lineno}
\usepackage{placeins}
\usepackage{xcolor}
\usepackage{lscape}

\newcommand\aj{{Astron. J. }}%
\newcommand\araa{{Annu. Rev. Astron. Astrophys. }}%
\newcommand\apj{{Astrophys. J. }}%
\newcommand\apjl{{Astrophys. J. Lett. }}%
\newcommand\apjs{{Astrophys. J. Suppl. }}%
\newcommand\aap{{Astron. Astrophys. }}%
\newcommand\icarus{{Icarus }}%
\newcommand\mnras{{Mon. Not. R. Astron. Soc. }}%
\newcommand\nar{{New A Rev. }}%
\newcommand\ssr{{Space~Sci.~Rev. }}%
\newcommand\nat{{Nature }}%
\newcommand\gca{{Geochim.~Cosmochim.~Acta }}%
\newcommand\grl{{Geophys.~Res.~Lett. }}%
\newcommand\jgr{{J.~Geophys.~Res. }}%
\theoremstyle{thmstylethree}%
\usepackage{xr}
\newcommand{\eg}{{\it e.g.}}
\newcommand{\al}{$^{26}$Al }
\newcommand{\alcomma}{$^{26}$Al, }
\newcommand{\alstop}{$^{26}$Al. }

\makeatletter
\newcommand*{\addFileDependency}[1]{% argument=file name and extension
  \typeout{(#1)}
  \@addtofilelist{#1}
  \IfFileExists{#1}{}{\typeout{No file #1.}}
}
\makeatother

\begin{document}

%\linenumbers

\title{\vspace{-1.5cm} Rapid formation of exoplanetesimals revealed by white dwarfs }

%%=============================================================%%
%% Prefix	-> \pfx{Dr}
%% GivenName	-> \fnm{Joergen W.}
%% Particle	-> Sec.pfx{van der} -> surname prefix
%% FamilyName	-> Sec.ur{Ploeg}
%% Suffix	-> Sec.fx{IV}
%% NatureName	-> \tanm{Poet Laureate} -> Title after name
%% Degrees	-> \dgr{MSc, PhD}
%% \author*[1,2]{\pfx{Dr} \fnm{Joergen W.} Sec.pfx{van der} Sec.ur{Ploeg} Sec.fx{IV} \tanm{Poet Laureate} 
%%                 \dgr{MSc, PhD}}\email{iauthor@gmail.com}
%%=============================================================%%

\author*[1]{\fnm{Amy} {Bonsor}}\email{abonsor@ast.cam.ac.uk}

\author[2,3]{\fnm{Tim} {Lichtenberg}}
\equalcont{These authors contributed equally to this work.}

\author[4,5]{\fnm{Joanna} {Dr{\c{a}}{\.z}kowska}}
\equalcont{These authors contributed equally to this work.}

\author[1]{\fnm{Andrew} M. {Buchan}}
\equalcont{These authors contributed equally to this work.}

\affil[1]{\orgdiv{Institute of Astronomy}, \orgname{University of Cambridge}, \orgaddress{{Madingley Rise}, \city{Cambridge}, \postcode{CB3 0HA}, \country{United Kingdom}}}

\affil[2]{\orgdiv{Kapteyn Astronomical Institute}, \orgname{University of Groningen}, \orgaddress{{Landleven 12}, \city{Groningen}, \postcode{9747 AD}, \country{Netherlands}}}

\affil[3]{\orgdiv{Atmospheric, Oceanic and Planetary Physics, Department of Physics}, \orgname{University of Oxford}, \orgaddress{{Parks Rd}, \city{Oxford}, \postcode{OX1 3PU}, \country{United Kingdom}}}

\affil[4]{\orgdiv{University Observatory, Faculty of Physics}, \orgname{Ludwig-Maximilians-Universität München}, \orgaddress{{Scheinerstraße 1}, \city{Munich}, \postcode{81679}, \country{Germany}}}

\affil[5]{\orgdiv{Max Planck Institute for Solar System Research}, \orgaddress{{Justus-von-Liebig-Weg 3}, \city{Göttingen}, \postcode{37077}, \country{Germany}}}

%%==================================%%
%% sample for unstructured abstract %%
%%==================================%%

\abstract{
% BACKGROUND (build tension)
The timing of formation for the first planetesimals determines the mode of planetary accretion and their geophysical and compositional evolution. Astronomical observations of circumstellar discs and Solar System geochronology provide evidence for planetesimal formation during molecular cloud collapse, much earlier than previously estimated. 
%However, planetesimals are invisible to disk observations, and disk sampling biases hinder directly connecting protoplanet growth with dust depletion.}
% OBJECTIVES/METHODS
Here, we present distinct observational evidence from white dwarf planetary systems for planetesimal formation occurring during the first few hundred thousand years after cloud collapse in exoplanetary systems.
% RESULTS
A significant fraction of white dwarfs have accreted planetary material rich in iron core or mantle material. In order for the exo-asteroids accreted by white dwarfs to form iron cores, substantial heating is required. By simulating planetesimal evolution and collisional evolution we show that the most likely heat source is short-lived radioactive nuclides such as \al ($t_{1/2} \sim 0.7$ Myr). 

% CONCLUSIONS
Core-rich materials in the atmospheres of white dwarfs, therefore, provide independent evidence for rapid planetesimal formation, concurrent with star formation.  }

\maketitle

\section{Main}\label{sec1}

\begin{figure}
\includegraphics[width=0.9\textwidth]{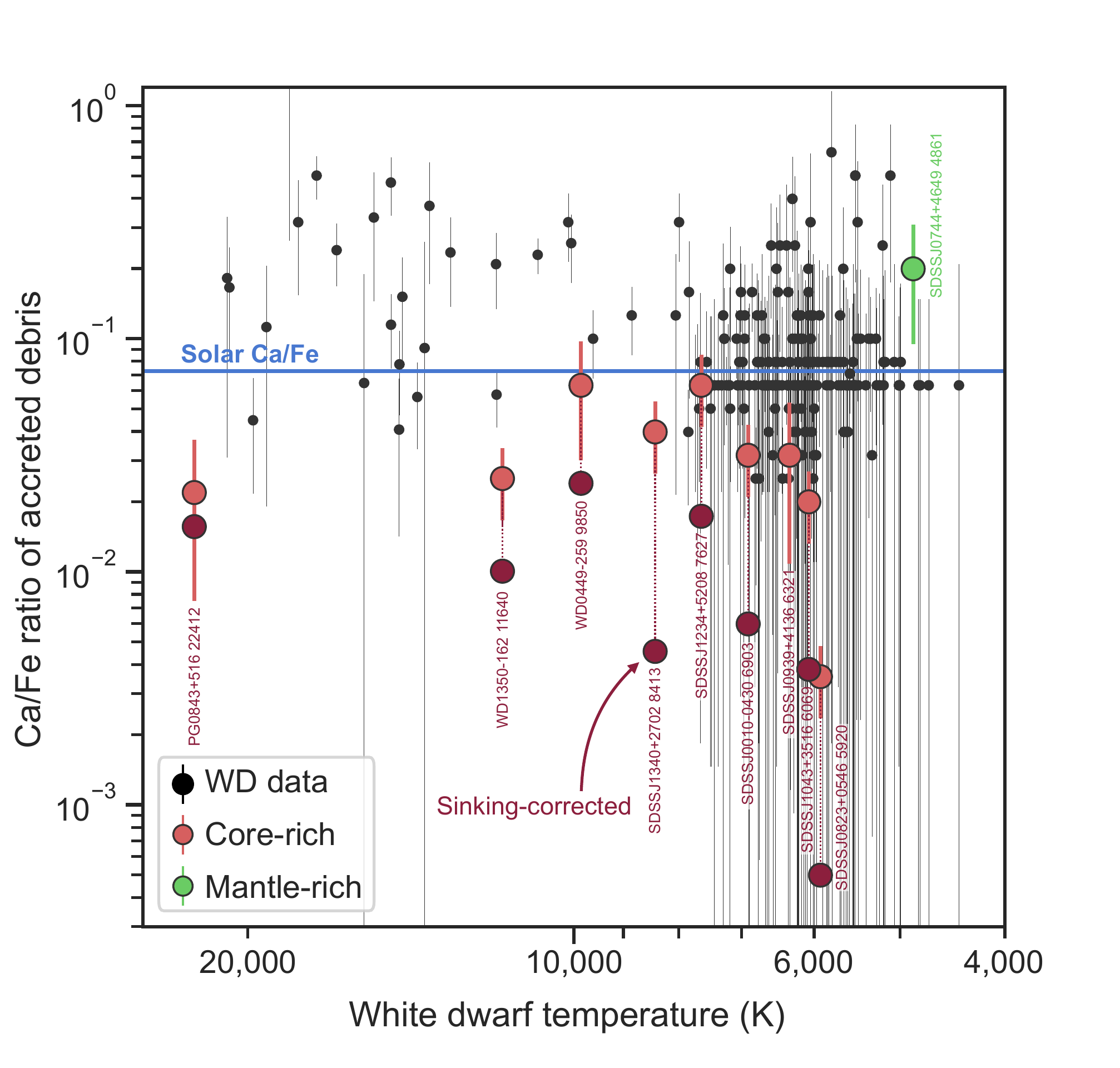}
 \caption{ {\bf Enrichment in Fe, Ni, and Cr relative to Ca, Mg, and Si of planetary materials accreted by white dwarfs suggest the accretion of core- or mantle-rich material.} Shown are the Ca/Fe ratios observed in a sample of 237 white dwarfs, alongside associated $1\sigma$ errors, as a function of white dwarf temperature. The large red circles indicate the 8 white dwarfs where a model in which core-rich material is accreted explains the observed abundances of all elements to $>3\sigma$ above a primitive model. In some cases the observed Ca/Fe is higher than the Ca/Fe in the accreted debris due to relative sinking, in which case the corrected abundances in the accreted material are plotted in dark red. SDSSJ0744+4649 shown in green has Ca/Fe$=0.2$ \cite{Hollands2017} and high Na, potentially related to the accretion of material from planetary lithosphere \cite{Harrison2021}. Models from \cite{Harrison2018, Harrison2021, Buchan2021}. The blue line indicates a solar Ca/Fe ratio. }
%\label{fig:wd_ncore}
\end{figure}

\begin{figure}
\includegraphics[width=0.9\textwidth]{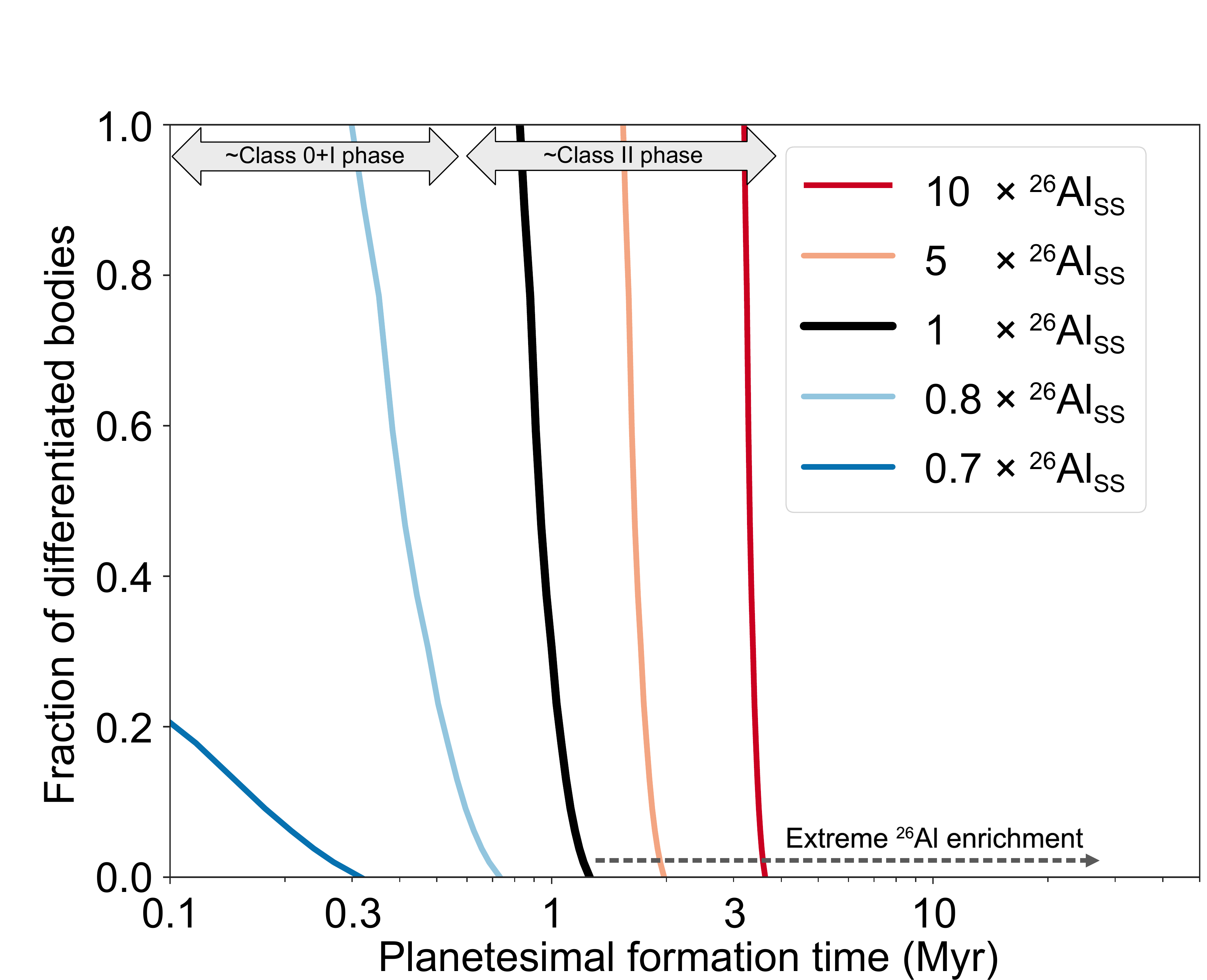}
 \caption{ {\bf Almost all planetesimals that undergo core-mantle differentiation form within the Class 0/I collapse phase in exoplanetary systems with plausible levels of \al enrichment. }Plotted is the fraction of planetary bodies likely to pollute white dwarfs (50--300 km in diameter) with sufficient \al to form an iron core \citep[][]{LichtenbergDrazkowska2021}, as a function of the time at which they formed. A size distribution of $n(D)dD \propto D^{-3.5}dD$ is assumed, and shown are a range of \al budgets, above and below Solar System levels {\bf ($^{26}$Al$_\mathrm{SS}=5.25 \times 10^{-5} \,\, ^{27}$Al$_\mathrm{SS}$ ) .} Few planetary systems have abundances significantly above solar \cite{Lugaro18,Lichtenberg2016,Kuffmeier2016,2021arXiv211109781F,Forbes21,2020A&A...644L...1R,2019ApJ...878..156C}.  }
%\label{fig:al26}
\end{figure}

\begin{figure}
\includegraphics[width=0.99\textwidth]{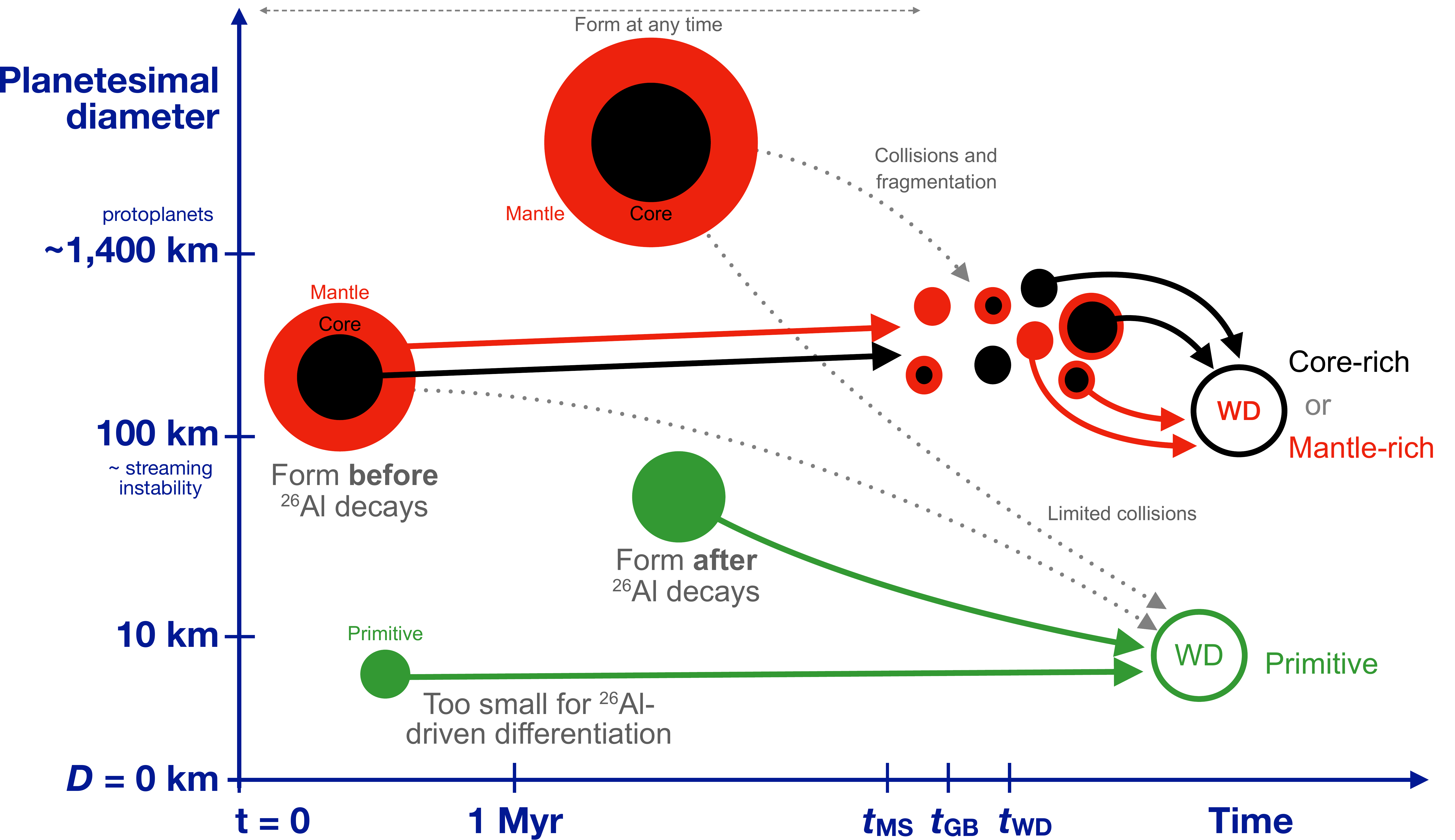}
 {\bf Fig. 3} {\bf The core- or mantle-rich materials in the atmospheres of white dwarfs are the collision fragments of planetesimals that formed earlier than $\sim$1 Myr, when large-scale melting was fueled by the decay of $^{26}$Al.} Alternatively, in the most massive, close-in, highly excited, planetesimal belts, catastrophic collisions between Pluto-sized bodies (anything with $D>1,400$ km) could supply most smaller planetesimals. Gravitational potential energy during accretion can fuel large-scale melting and core formation in these large bodies, such that almost all planetary bodies in the belt are the collision fragments of core--mantle differentiated bodies. $t_{\rm MS}$, $t_{\rm GB}$ and $t_{\rm WD}$ refer to the star's main-sequence, giant branch lifetimes and the start of the white dwarf phase. % in \S\ref{sec:collisionmodel}).}
\label{fig:cartoon}
\end{figure}

\begin{figure}
\includegraphics[width=0.98\textwidth]{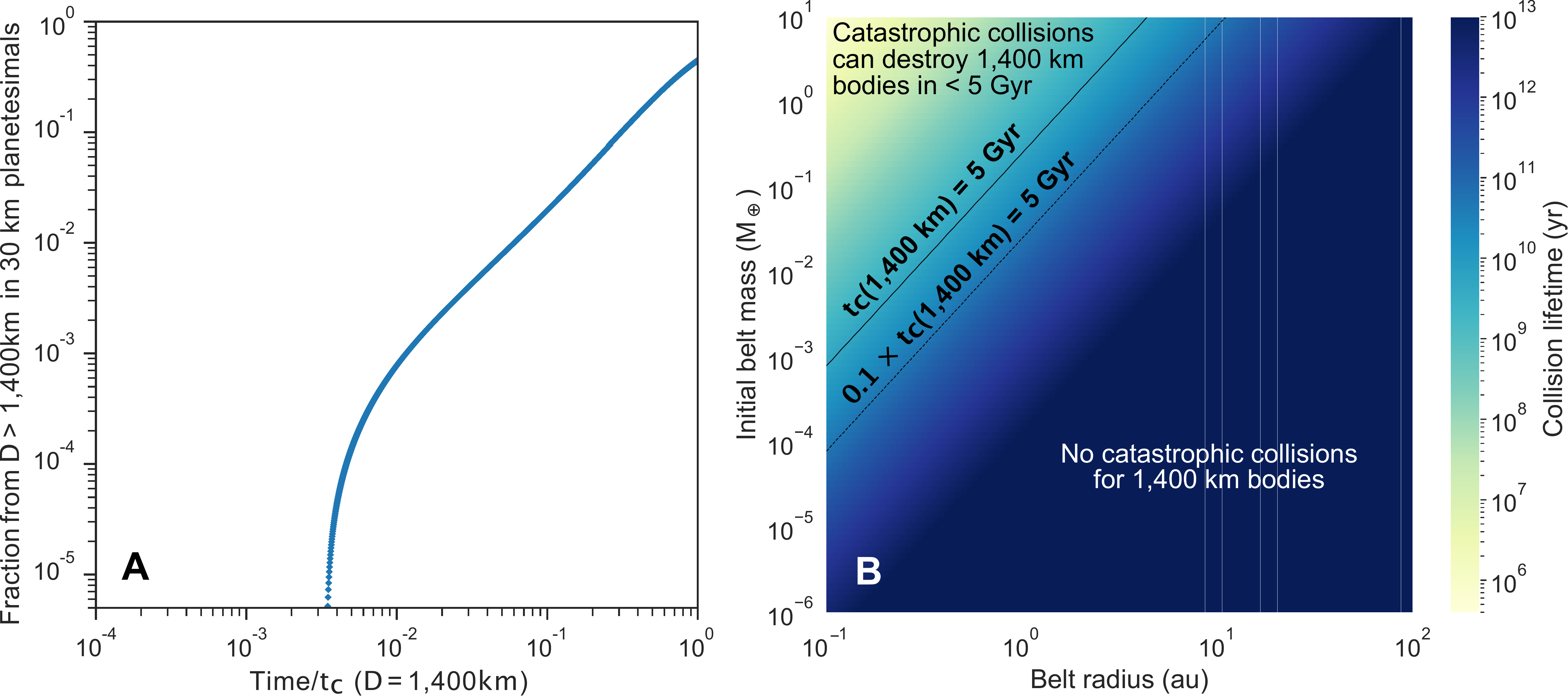}
 {\bf Fig. 4} {\bf Pluto-sized bodies can be the source of core-rich planetesimal debris only in rare ($<$1\%) white dwarf systems with massive, close-in planetesimal belts.} \textbf{(A)} The fraction of 30 km debris that are fragments of Pluto-sized core--mantle differentiated planetesimals ($D>1,400$km) (in units of the collision lifetime, Eq.~\ref{eq:t_c}) for a belt at 1au, with average particle eccentricity $<e>=0.1$ and initial mass of $100M_\oplus$ in particles between $100\mu$m and 5,000km. \textbf{(B)} Approximation to the collision lifetime as a function of the initial mass in the planetesimal belt in bodies between 100$\mu$m and 1,400km in diameter and the belt radius (Eq. ~\ref{eq:t_c}). A collision lifetime of 5 Gyr is shown by the solid black line and 10\% of this collision lifetime by the dashed black line. Less than a percent of debris discs, those with very massive, close-in planetesimal belts, that lie in the top left-hand corner above the solid line, will have catastrophic collisions of Plutos ($D>1,400$ km bodies) supplying material to the smaller planetesimals that might pollute white dwarfs, based on typical properties of observed debris discs. This is too low to explain the $4\%$ (Sample One, \ref{sec:sampletwo}) to $>13\%$ (Sample Two, \ref{sec:sampleone}) of white dwarf pollutants that accreted fragments of core-mantle differentiated bodies. 
%\label{fig:time_fracdiff}
\end{figure}

The timing and locations of planetesimal formation are crucial to our understanding of planet formation. If we are to form larger planets; gas giants or terrestrial planets, we must first form their building blocks; planetesimals. The meteorite record provides strong evidence that planetesimal formation in the Solar System spanned a wide range of ages, with magmatic iron meteorites dating $<$1 Myr after the formation of Ca-Al-rich meteoritic inclusions (CAIs, the oldest known solids formed in the Solar System) \cite{2006E&PSL.241..530S,2014Sci...344.1150K}, whilst carbonaceous chondrite meteorites record formation times extending to $\approx$5 Myrs after CAIs \cite{2020SSRv..216...55K}.  The key question for understanding the growth mechanism of planets such as Jupiter is whether planetesimals form sufficiently early to allow time for the accretion of larger protoplanets prior to the end of the circumstellar disc, whose lifetimes are typically several Myrs \cite{Fedele2010}. Without a knowledge of the timing of CAI formation, it is difficult to pin down whether planetesimal formation started in the Solar System during the collapse phase, traced observationally by Class 0/I discs, while the protostar is still accreting from the surrounding molecular cloud, or later, in Class II discs that are spatially isolated from their star-forming environments.% \citep{Williams2011}.  
 
Traditional planet formation models start with fully-fledged, Class II discs, assuming that all the solids are in the form of dust and the dust evolution only starts at the beginning of the Class II phase. % \cite{Weidenschilling1980}. %., Brauer2008}. 
%Such models are successful in ... however, they struggle to form gas giant planets before the gas disc dissipates. 
Observationally, Class II discs do not contain sufficient material in dust to form the observed population of exoplanets \cite{Najita2014,Tychoniec2020}. Observed substructures in very young circumstellar discs \cite{Sheehan2018, SeguraCox2020} may indicate the presence of over-densities where planet formation may already be underway during the Class 0/I stage \cite{Stammler2019, Carrera2021}, although these structures can alternatively be explained by disc instabilities or condensation fronts \cite{Flock2015, Zhang2015}. Probing these discs with the Atacama Large Millimeter/submillimeter Array (ALMA) reveals the mass in mm/sub-mm grains (dust) as probed by its thermal emission, but planetesimals and larger protoplanets are invisible at ALMA wavelengths. Thus, the main observational way to probe the growth of planetesimals is to search for trends in dust depletion with disc stage, which are complicated by correlations between disc structure, size and disc stage, as well as observational biases in the disc and exoplanet populations \cite{vanderMarel2021, Mulders2021ApJ}. Therefore, further evidence regarding the timing of planetesimal formation is required to test the main channels and timescales of planetary growth.

In this work, we present distinct observational evidence that planetesimal formation commenced early in a significant fraction of exoplanetary systems. This evidence comes from white dwarfs that have accreted planetary material. Fragments of planetary bodies from a surviving outer planetary system show up in the spectra of an otherwise clean (hydrogen/helium only) white dwarf \cite{JuraYoung2014, Farihi_review}. From these observations the composition, notably ratios of key elements such as Si, Mg, Fe, O, Ca, C, Cr, or Ni in the planetary material can be found. Elements heavier than helium should sink out of sight on timescales of days to millions of years, depending on the white dwarf temperature, surface gravity and atmospheric composition \citep{Fontaine1979, Koester2009}. Thus, the observed material must have arrived recently. Planetary material is found in a significant proportion of white dwarfs \citep[30-50\%,][]{Zuckerman2010, Koester2014}, with observations able to detect relatively small amounts of material (equivalent to km-sized asteroids). For most white dwarfs, the observed abundances are consistent with the accretion of primitive rocky material, but for some white dwarfs, there is an over- or underabundance of core affine (siderophile) species such as Fe, Cr, Ni  relative to mantle affine (lithophile) species, such as Mg, Si, which is best explained by metal-silicate partitioning that occurs during the formation of an iron core \citep{Melis2011, Gaensicke2012, Wilson2015}. These white dwarfs have accreted a fragment of the metal core or silicate mantle of a chemically differentiated planetary body \citep{JuraYoung2014}.
%% Note that the two objects with high Ca/Fe - one is declining and one is crust-rich [0.19952623 0.06309573]
%39     SDSSJ0744+4649
%116    SDSSJ1234+5208

Observationally, a significant fraction of white dwarfs with planetary material in their atmospheres have accreted core- or mantle-rich material. Conservatively, in a sample of more than two hundred white dwarfs, based primarily on Ca, Fe and Mg abundances, 4\% are best explained (to $>3\sigma$) by the accretion of core-rich material (Sample One: \ref{sec:sampletwo}). When more elements are detected, more information regarding the planetary material can be deduced. In the 54 white dwarfs with more than 5 elements detected considered here (Sample Two: \ref{sec:sampleone}), $7$\% were best explained (to $>3\sigma$) by a model that invokes core--mantle differentiation \citep{Buchan2021} noting, however, that this sample was not selected in a uniform manner. The models used \cite{Harrison2018, Harrison2021, Buchan2021} place stringent conditions on invoking core--mantle differentiation, take into account the abundances of all elements observed in each system, account for relative sinking, as well as volatile depletion and potential variations in the initial composition of the planet-forming material (\ref{sec:models}). Only those fragments with extremely core- or mantle-rich compositions will be identified, although we caveat here that additional processes such as impact melting, the suggested origin to low Ca/Fe in CB chondrites \citep{Krot2005} are not included in the current models. The Ca/Fe ratios of the planetary material accreted by the 237 white dwarfs in both samples considered are shown in Fig.~1 as a function of white dwarf temperature, with the large circles indicating those objects with a $>3\sigma$ requirement for core-rich material, noting that the model does not identify many mantle-rich fragments (to $>3\sigma$) due to a degeneracy between mantle-rich compositions and the depletion of moderately volatile elements.

The segregation of material between the iron-rich core and silicate mantle requires large-scale melting. If the white dwarfs accreted exo-asteroids, the most likely source of energy to fuel the large-scale melting is the decay of short-lived radioactive nuclides \citep{Jura2013}. As seen in the Solar System \citep{1977GeoRL...4..299W}, $^{26}$Al fuels large-scale melting, with alternate species such as $^{60}$Fe largely absent from the solar disc \citep{Tang2012}. Here, we show that it is unlikely that the white dwarfs accrete minor planets, nor the collision fragments of minor planets, where the large-scale melting could have been fueled by gravitational potential energy. \al has a half-life of 0.717 Myr and its heating potential dwindles rapidly after $\sim$1--2 half-lives. For planetesimals to contain sufficient \alcomma they must form early, within the first Myr of the evolution of the planetary system, when sufficient \al for melting and large-scale differentiation still abounds.   

% $^{26}$Al

The distribution of short-lived radioactive nuclides across exoplanetary systems is unknown \citep{Lugaro18}, with end-member inferences ranging from a small fraction of exoplanetary systems (a few percent) \citep[\eg][]{Gounelle2015} to a significant fraction, potentially the majority of planetary systems \citep[\eg][]{Young2014} featuring Solar System-like abundances. Most works, however, suggest that few systems have significantly higher abundances of \al than the Solar System \citep{Lichtenberg2016,Kuffmeier2016,2019ApJ...878..156C,2021arXiv211109781F}, which is supported by observational evidence from individual star-forming regions \citep{Forbes21,2020A&A...644L...1R}. Depending on when planetesimal formation occurs, this means that for some exoplanetary systems, with high initial budget of short-lived radioactive nuclides, a large fraction of planetesimals will form an iron core. For other exoplanetary systems, with lower levels of enrichment, only the small fraction of planetesimals that form early segregate to a differentiated mantle/core structure. 
Fig.~2 illustrates this point, by showing the fraction of planetesimals likely to pollute a white dwarf (chosen to be between 50 and 300 km in diameter, approximately the birth size range produced by the streaming instability) that contain sufficient \al to form an iron core, as a function of the time at which they formed and the initial abundance of \al in the system. This is calculated based on the bodies reaching a mean internal temperature above which planetesimals can experience core-mantle differentiation by percolation of metal-sulfide liquids using the models of \cite{LichtenbergDrazkowska2021} and assuming a size distribution in planetesimals of $n(D)dD \propto D^{-7/2} dD$. Almost all planetesimals that form earlier than $\sim 1$Myr form an iron core, whilst almost no bodies that form later than a few Myr contain sufficient \al to lead to large-scale melting. Even at 5 times higher abundances of \al than solar, only a few bodies that form later than 2 Myr can form iron cores. Varying the initial size distribution and upper/lower bounds of the planetesimal population within plausible limits only marginally affect these overall conclusions.

Thus, if \al fuels the large-scale melting, the observations of core- or mantle-rich material accreted by white dwarfs requires the early formation of planetesimals in exoplanetary systems, most likely within the first Myr after the injection of \alstop With \al injection at (or before) the start of the collapse of the molecular cloud \citep{2019ApJ...878..156C,Forbes21,2020A&A...644L...1R,Lugaro18}, the white dwarf observations thus provide evidence that planetesimal formation occurred already during the Class 0/I phase. %Although the duration of this collapse phase is debated, observationally most Class I objects are younger, but some are still found at 0.5 Myrs \citep{SeguraCox2020} and models suggest disc formation takes up to 0.7 Myrs \citep{dd2018}.

A schematic illustrating of the proposed scenario is shown in Fig.~3. Planetesimals that form early in systems with a sufficient budget of short-lived radioactive nuclides will undergo large-scale melting and form an iron core, as occurred for iron meteorite parent bodies in our Solar System. Leftover planetesimals, not incorporated into planets, form collisional belts, as witnessed by observations of debris discs \citep{Hughes_revew}. Violent collisions can produce core- or mantle-rich fragments \citep{Marcus2010, Carter2015}. 
These fragments evolve in planetesimal belts, those of which are exterior to a few au, survive dramatic phases of evolution as their host stars become giants and lose their outer envelopes to start the white dwarf cooling phase. Scattering by planets, or other dynamical instabilities following stellar mass loss, can lead to some of these fragments being accreted by white dwarfs \citep{DebesSigurdsson}, where their core- or mantle-rich compositions show up in the atmosphere. Those planetary bodies that formed after \al decayed, undergo the same collisional evolution, scattering and accretion, but show up as {\it primitive} compositions in the atmosphere of the white dwarf. Thus, if the parents of the white dwarf pollutants are asteroids, the presence of core or mantle material is evidence for their formation within the first few hundred thousand years of cloud collapse. 

Alternatively, as indicated by the dotted lines on Fig.~3, planetary bodies larger than about 1,400 km may form an iron core without the need for \alstop For such large bodies sufficient gravitational potential energy is available during formation to lead to large-scale melting \citep{Elkins-Tanton2011} (\ref{sec:large}). Moons or even terrestrial planets undergo magma ocean phases and form iron cores due to this gravitational potential energy, but are rare (by number) relative to asteroids. Whilst dynamical mechanisms exist for the liberation of exo-moons or the direct scattering of planets onto white dwarfs, these pathways seldom occur \citep{Payne2016, Veras_twoplanet_2013}. This is in stark contrast to the ubiquitous nature of white dwarf pollution, with 30-50\% of white dwarfs having planetary material in their atmospheres \citep{Zuckerman2010, Koester2014}, pointing towards the accretion of moons/planets as an unlikely pathway for most pollution of white dwarfs. Nor, are the core-rich systems outliers with higher than average accretion rates. Additionally, the observed masses and inferred accretion rates for all, but a handful of cool white dwarfs, are asteroidal masses (or smaller) \cite{Veras_review}. In order to accrete an Earth mass of material, accretion would need to be moderated at low accretion rates and continue on Gyr (or longer) timescales since there are no observed accretion rates higher than $\sim10^{11}$gs$^{-1}$ \citep{Farihi2012}.

Theoretically, the largest planetary bodies within a planetesimal belt may in principle form iron cores without the need for \alstop The existence of such large bodies within exoplanetesimal belts is debated, due to the rapid decrease in the brightness of discs with time, which would not occur if collisions between large bodies were replenishing the small dust \citep{Krivov_Wyatt2021}. If a population of {\it Plutos} exist, their catastrophic collisions can dominate the mass budget of massive, close-in (less than a few au) planetesimal belts (\ref{sec:collisionmodel} \citep{wyattreview}). In this scenario, most small planetary bodies are the collision fragments of Plutos. Thus, the 10-100 km asteroids polluting white dwarfs would likely show up with core- or mantle-rich compositions. The fraction of 30 km planetesimals that are fragments of Plutos ($D>1,400$ km) is shown in the left-hand panel of Fig.~4 as a function of time. On timescales less than 10\% the collision lifetime (0.1 $t_c$ ($D$=1,400 km), Eq.~\ref{eq:t_c}), less than a percent of the 30 km planetesimals plausibly polluting white dwarfs would be collision fragments of core--mantle differentiated, $D_*=1,400$ km, planetesimals. %Noting that as the collisional cascade is self similar, the plot is created for the less computationally intensive simulations considering the fraction of planetesimals that are fragments of bodies larger than diameter, $D_*=1,2,5$ km, but that the plot would look equivalent for any plausible $D_*$. The differences between $D_*=1,2,5$ km arise {\bf as the collisional casacade is not a perfect power-law, when a size cut-off for small dust grains is considered}.  
Thus, the proposed scenario can only occur in planetesimal belts where collisional evolution has proceeded for longer than the collisional lifetime of Plutos. The right-hand panel of Fig.~4 shows that only very massive, close-in planetesimal belts have a sufficiently short collision lifetime for Plutos, approximately a percent of planetesimal belts, based on the distribution of planetesimal belt properties that fits current observational samples \cite{wyatt07}. Additionally, only a small fraction (on the order of 10\%) of planetesimals in such systems would have compositions sufficiently core- or mantle-rich to be detected.

The white dwarf observations suggest that enrichment by \al is common across exoplanetary systems. Large-scale melting fueled by gravitational potential energy in Plutos or larger bodies is only likely to account for a tiny ($\lt 0.1$\%) fraction of white dwarf pollutants. Apart from the direct consequences for core-mantle differentiation, the common enrichment of exoplanetary systems by \al has far-reaching implications for the volatile budgets of rocky planets acquired during formation. Planetary bodies that form exterior to ice-lines loose their volatiles due to heating from \alcomma introducing a disconnect between ice-lines and the volatile content of planets \citep{LichtenbergDrazkowska2021,2021ApJ...913L..20L}. As the abundance and fractionation of highly volatile elements on rocky planets is key to their long-term climate \cite{2021arXiv211204663W}, our findings highlight the influence of short-lived radioactive nuclides on the surface conditions and frequency of potentially temperate, Earth-like exoplanets. The need for enhanced abundances of \al to explain core- or mantle-rich white dwarf spectra provides distinct evidence for the early formation of planetesimals in exoplanetary systems contemporaneously with star formation. Rapid planetesimal formation offers an explanation for the difference in mass budgets between Class 0, I and II discs \cite{Tychoniec2020}. Our findings point to the growth of large, $>$10 km-sized planetesimals, potentially even planetary cores, rather than just the coagulation of pebbles. The earlier planetary cores form, the more likely they are to grow to the pebble isolation mass and the more likely giant planet formation is to occur early-on \cite{Drazkowska2021}, which can provide an explanation for substructures commonly observed with ALMA. A new picture is emerging of star and planet formation starting concurrently, with large planetary bodies forming and geophysically evolving already during the collapse of the planet-forming disc, traditionally associated with Class 0/I systems.

\section{Methods}
In order to determine how frequently the planetary bodies accreted by white dwarfs underwent large-scale melting and differentiated internally, core and mantle-rich compositions were identified by analysing the abundances observed in two distinct samples of polluted white dwarfs. The first is selected for outcome ($>5$ elements detected) and contains predominantly white dwarfs with high quality data, whilst the second contains only DZs, observed and analysed in the same manner, based on their SDSS spectra. The following sections describe the models used to explain the observed abundances and the two white dwarf samples considered here.
\subsection{Models to explain the abundances observed in the atmospheres of white dwarfs}
\label{sec:models}

 The white dwarfs considered here all have spectra in the optical and/or UV, with abundances for a number of metals species in the hydrogen or helium atmosphere previously presented in the literature.  The most likely explanation for the observed abundances is found  using Bayesian models presented in \citep{Harrison2018, Harrison2021, Buchan2021} (\url{https://github.com/andrewmbuchan4/PyllutedWD_Public}). The results for most white dwarfs considered were presented previously in \cite{Harrison2021, Buchan2021}, with those analysed specifically for this paper detailed in Extended Data Table~1. These models consider all the elements that have been detected, alongside upper limits where available. These models do not take into account S, Sc, Cu, Co, V, P, Mn, Ga, Ge, K, Li or Be. The potential that the observed abundances are altered from those in the accreted planetary material due to relative sinking is considered. A range of initial conditions for the planetary material are considered, with the compositions of nearby stars \citep{Brewer2016} used as a proxy for this range. The abundances in the planetary material can be altered due to loss of volatiles, which for the simplest scenario is just the loss of water to make rocky asteroids. However, all elements, including moderate volatiles such as Na, are considered and this loss of volatiles is modelled as the incomplete condensation of the nebula gas in chemical equilibrium. The white dwarf is then allowed to accrete a fragment of a larger planetary body with the core mass fraction being a free parameter. In other words, the white dwarf could accrete a chunk of the iron core (core mass fraction $= 1$) or a chunk of silicate mantle (core mass fraction = 0), or a chunk of predominantly core material with some mantle remaining (e.g. core mass fraction = 0.9) and so on. The composition of the core and mantle material is allowed to vary depending on the pressure and oxygen fugacity conditions under which the planetary body formed its iron core, using metal-silicate partitioning parameterised according to \cite{Fischer2015, Corgne2008, WadeWood2005, Wood2008, Cottrell2009, Siebert2012}.   

\subsection{White Dwarf Observations}

\subsubsection{Sample One: Cool DZs from SDSS \cite{Hollands2017, Hollands2018}}
\label{sec:sampletwo}
202 cool white dwarfs with only metal features (DZ) were selected from their SDSS spectra with detections of at least Mg, Fe and Ca from \cite{Hollands2017, Hollands2018}. We note here that magnetic or unresolved binary white dwarfs were not included in the sample and that updated abundances from \cite{Blouin2020, Harrison2021} were used. The spectra have relatively low S/N compared to Sample Two targets and thus, fewer elements are detected and the uncertainties are larger. Those white dwarfs in this sample where more than 5 elements were detected are also included in Sample Two. These white dwarfs were predominantly selected due to their colours in SDSS (u-g) (g-r) space, where the large absorption features due to the presence of metals in these white dwarf spectra moves the white dwarfs from  above the main-sequence to below the main-sequence. This selection function may bias the sample towards white dwarfs with high Ca abundances, however, the requirement that Fe and Mg must also be detected, means that the distribution of Ca/Fe in the sample is only slightly skewed to high Ca/Fe \citep{Bonsor2020}. \cite{Harrison2021} analyse this sample of white dwarfs in detail and find crucially that mantle-rich fragments are harder to identify due to a degeneracy with sinking and volatile depletion. \cite{Harrison2021} identify 7/202 (4\%) white dwarfs where the accretion of core-rich material is required to $>3\sigma$ over the accretion of primitive material. We note here that \cite{Harrison2021} incorrectly stated 8 white dwarfs were best explained by the accretion of core-rich material, when 8 white dwarfs were best explained by the accretion of core--mantle differentiated material.  One object (SDSSJ0744+4649) is identified, where the Ca, Fe, Mg abundances suggest an enhancement of Ca and Mg relative to Fe, as seen in planetary mantles, with the enhanced Na indicating that this cannot be volatile depletion \citep{Harrison2021}. The full details of the sample are presented in the Supplementary Information of \cite{Harrison2021}.

\subsubsection{Sample Two: white dwarfs with more than 5 elements detected}
\label{sec:sampleone}
 54 white dwarfs were selected from the literature with abundances of more than 5 elements, including Fe. These white dwarfs tend to be the most highly polluted, the brightest stars and the most studied objects.  19 of these white dwarfs were also included in Sample One. Most have high resolution spectra, potentially from multiple instruments. By necessity, however, the selection of the sample is observationally biased, with many observations tending to target those objects that are easiest to measure. The atmospheric abundances were analysed using the model presented in \cite{Buchan2021} which updates the models of \cite{Harrison2021} by modelling core--mantle differentiation without any assumption of Earth-like material. Whilst the most likely explanation (highest Bayesian evidence) for the observed abundances includes core-mantle differentiation for a third of the sample (19/54), the abundances are consistent, within the errors, for most white dwarfs with the accretion of primitive material, whose abundances are only altered by volatile loss, sinking in the white dwarf atmosphere and the potential small variation in the composition of the initial planet forming material. For an additional 3 systems (NLTT43806, LHS 2534 and SDSSJ0744+4649), previous work has suggested the accretion of crust-rich material to explain the abundances \citep{Hollands2021, Zuckerman2011, Harrison2021}. The model used here does not account for crustal differentiation. 

In identifying those white dwarfs that potentially accreted core or mantle-rich fragments of larger planetary bodies, the relatively large uncertainties on the atmospheric abundances, as well as the unknown time since accretion started, which determines the relative sinking of elements, play a significant role. In many cases the Bayesian models finds the highest evidence for a model which invokes core-rich material. This is indicated by the Bayes factor, which \cite{Harrison2021} and \cite{Buchan2021} convert to a sigma significance \citep{Sellke2001} using Eq.10 of \cite{Harrison2021}. We focus here on those systems where $\sigma_{\rm diff}>3$, although noting that core-rich material may well be the true explanation for systems with $\sigma_{\rm diff}<3$. Core-mantle differentiation is required ($>3\sigma$) to explain the abundances in 4/54 (7\%) of systems (PG 0843+516, SDSSJ1043+3516, WD0449-259, WD1350-162), although noting that in Sample One, for the two systems SDSSJ0939+4136 and SDSSJ1234+5208 the Earth-like differentiation models of \cite{Harrison2021} increased the significance to which core--mantle differentiation was invoked from slightly below to over 3. Including the 3 crust-rich systems, at least 7/54 (13\%) underwent large-scale melting and plausibly a significantly higher fraction. The sample is slightly different from that presented in \cite{Buchan2021}, now including 19 additional objects with more than 5 elements detected, but which did not include Ni, Cr or Si, as required by \cite{Buchan2021}, whilst not including objects with $<5$ elements detected. However, the analysis is identical to that performed by \cite{Buchan2021}, which updates the model of \cite{Harrison2021} to allow for core-mantle differentiation in systems with arbitrary, rather than Earth-like compositions.

The full list of white dwarfs in the sample is presented in Extended Data Table~1, alongside the atmospheric abundances used in this work in Extended Data Table~2 and the most likely model parameters, as determined by the Bayesian models are presented in Extended Data Table~3. We note here that the model has been updated since \cite{Harrison2018}, also including updated sinking timescales, as well including stricter criterion for where the accretion of core--mantle differentiated material is required to explain the observed abundances. We note here that a discrepancy exists between abundances determined from UV and optical data (see \cite{Xu2019} for more details). For a number of white dwarfs where conflicting abundances exist, a consistent set of abundances from the UV was used and is noted in Extended Data Table~2.

\subsection{Gravitational potential energy as a driver of core-mantle differentiation.}
\label{sec:large}

During the formation of the largest planetesimals, or indeed moons or terrestrial planets, there is sufficient gravitational potential energy available that when this is converted to heat, large-scale melting can occur. In order to estimate how large a planetesimal must be for there to be sufficient gravitational potential energy, the energy deposited in a body by the accretion of smaller objects, per unit mass, is considered to be $E~\sim \frac{h}{2} (v_{\rm esc}^2 + v_{\rm rel}^2)$, where $h$ is the fraction of the energy deposited as heat, rather than re-radiated, $v_{\rm esc}$ the escape velocity of particles from a planetesimal of mass $M$ and radius $R$, and $v_{\rm rel}$ the relative velocity between the particles and the growing planetesimal, following \cite{Elkins-Tanton2011}. Given that the relative velocity of most particles is approximately the escape velocity, this becomes $E~\sim \frac{hGM^2}{R}$, which for spherical planetesimals of uniform density is approximately, $E= \frac{hG \rho R^2 4 \pi}{3}$. The energy required to raise the temperature from typical temperatures at the mid-plane of proto-planetary discs (around 700K) to the temperatures required for large-scale melting ($\sim 1,200$K), assuming the specific heat capacity of the body is around that for silicates ($C_p= 800$J kg$^{-1}$ K$^{-1}$) is $4\times 10^5 $J kg$^{-1}$ \citep{Elkins-Tanton2011}. Using a conservative $h=0.8$ and a density of 3 g cm$^{-3}$ a planetesimal of radius $>700$ km (diameter $>1,400$ km) can become differentiated by gravitational energy alone.

\subsection{Collisional evolution of planetesimal belts: could most planetesimals be fragments of Plutos?}  
\label{sec:collisionmodel}

One route to get core- or mantle-rich pollutants into the atmosphere of white dwarfs is to scatter in asteroids (10-300 km in size) that are themselves fragments of Plutos ($D>1,400$ km), bodies large enough to form an iron core without the need for heating from short-lived radioactive nuclides (see \ref{sec:large}). These bodies can form at any time (Fig.~3). If there are sufficient collisions in a planetesimal belt, the Plutos can reach collisional equilibrium and fragments of these large bodies will feed the population of smaller bodies in the belt. We present models for the collisional evolution of planetesimal belts that determine the fraction of asteroids ($D=10-300$ km) that are fragments of Plutos ($D>1,400$ km) as a function of time. In these systems, core- or mantle-rich fragments could be accreted by white dwarfs from planetesimals that formed at any epoch. We find that this is a rare pathway to white dwarf pollution. The simulations show that before smaller bodies are likely to be fragments of a larger body of a given size, $D_{\rm eq}$, those bodies must reach (or almost) reach collisional equilibrium, or in other words a time, $t_c(D_{\rm eq})$ (Eq.~\ref{eq:t_c}), must pass. As it takes a long time for Plutos to reach collisional equilibrium, this only occurs in the most massive, close-in planetesimal belts, of which too few exist for them to be the likely source of many white dwarf pollutants.

\subsubsection{Collision Model}
The model traces the collisional evolution of a planetesimal belt with time. The mass in the belt is split into logarithmically spaced bins and the origin of the mass in each size bin is traced as a function of time. In other words, the aim is to answer the question of whether most white dwarf pollutants (of size {\it e.g.} 30 km) are collision fragments of larger bodies, in particular bodies larger than $>1,400$ km.

The model for collisional evolution is based on \cite{wyatt11}, presented in detail in Bonsor et al, 2023, in prep. Here
we consider {\it solids only} and catastrophic collisions only. We consider the belt
to be a single annulus that contains particles from size $M_{\rm min}$ up to $M_{\rm max}$, or equivalently from
diameter $D_{\rm min}$ up to
diameter, $D_{\rm max}$, where spherical particles of constant density are assumed, such that particles in the $k$th bin of diameter, $D_k$, have a mass, $M_k = \frac{ \pi D_k^3}{6}$ with a size distribution: 
\begin{linenomath}
\begin{equation}
 n(M)\, dM\propto M^{-\alpha} dM
%%n(D) dD \propto D^{-\gamma} dD.
\label{eq:n_m}
\end{equation}
\end{linenomath}
We assume a standard, infinite
collisional cascade \citep{Dohnanyi, wyattreview}, with power law
index of $\alpha = 0.83$, or equivalently for diameter $q=3.5 = 3\alpha +1$.  
The size distribution is split into bins of equal width in log space ($\delta$), labelled
by their mass, $M_k$. The spacing, $\delta$,  is assumed to be small, such that $\frac{M_{k+1}}{M_k}=1-\delta$. At every
time-step, we calculate the rate at which each bin gains and loses
mass. We assign a fractional origin of material in each bin from every
other larger mass bin in the system. At each time-step, this fractional
origin of material is updated, taking into account the origin of the mass gained and lost
in each mass bin, as well as the mass that stays in this bin from previous time-steps.

In order to trace the collisional evolution of the material between size bins, a threshold is defined,
such that the smallest particle that can destroy a body of size $M_k$ is
given by: 
\begin{linenomath}
\begin{equation}
M_{ck}= \left ( \frac{2 Q_D^*}{v_{\rm rel}^2} \right) M_k
\label{eq:mck}
\end{equation}
\end{linenomath}
where $v_{\rm rel}$ is the relative velocity in collisions, $Q_D^*$ is
is the specific incident energy required to cause a catastrophic
collision, or the dispersal threshold. The ratio of the smallest size
that can destroy a body to its size is given by
$X_c=\frac{M_{ck}}{M_k}$.  We assume a power-law form for the dispersal threshold, following work on collision outcomes by \cite{benzaphaug, Durda98}, such that : 
\begin{linenomath}
\begin{equation}
Q_D^*=Q_a \left(\frac{D}{m}\right)^{-a} + Q_b \left(\frac{D}{m}\right)^b,
\label{eq:qdstar}
\end{equation} 
\end{linenomath}
where $a$ and $b$ are both positive constants related to the planetesimal's material and gravitational strength, respectively and $D/m$ is the planetesimal diameter in metres. Following \cite{wyatt11} 
we take $Q_a= 620$ Jkg$^{-1}$,   
$a = 0.3$, $Q_b=5.6 \times 10^{-3}$Jkg$^{-1}$
and $b = 1.5$. 
The rate of catastrophic collisions in the $k$th bin is given by, $R_k^{c}$ is given by: 
\begin{linenomath}
\begin{equation}
R_k^{c}= \Sigma_{i=1}^{i_{ck}} \frac{n_i}{4} (D_k + D_i)^2 P_{ik},
\end{equation}
\end{linenomath}
where $n_i$ is the number of colliders in the $i$th bin and $P_{ik}$
is the intrinsic collision probability,  $P_{ik}=\frac{\pi
  v_{rel}}{V}$, where $V$ is the volume through which the
planetesimals, of mass $M_k$ are moving. $i_{ck}$ refers to the smallest impactors
that can cause catastrophic destruction, of mass $M_{\rm ck}$ (Eq.~\ref{eq:mck}). 

 We consider that mass is conserved such that the total mass in each bin, ${m}_{{\rm s},k}$ is governed by the following equations:
 \begin{linenomath}
\begin{equation}
{\dot m}_{{\rm s},k} =  \dot{m}_{{\rm s},k}^{+c} - \dot{m}_{{\rm s},k}^{-c} 
\label{eq:massconservation}
\end{equation}
\end{linenomath}
where $\dot{m}_{{\rm s},k}^{-c}$ is the rate at which the total mass
in the $k$th bin is lost to catastrophic collisions, given by :
\begin{linenomath}
\begin{equation}
\dot{m}_{{\rm s},k}^{-c}= m_{{\rm s},k} R_k^{c}, 
\label{eq:mloss}
\end{equation}
\end{linenomath}

and $\dot{m}_{{\rm s},k}^{+c}$ is the rate at which the mass in solids is
gained from catastrophic collisions of larger bodies, given by:
\begin{linenomath}
\begin{equation}
\dot{m}_{{\rm s},k}^{+c}= \Sigma^{i_{mk}}_{i=1} F(k-i) \; \dot{m}_{{\rm s},i}^{-c} , 
\label{eq:mgain}
\end{equation} % is this the correct sum - surely only fragments that are 2mk or greater can contribute mass to the kth bin? 
\end{linenomath}

where $F(k-i)$ is the fraction of the mass leaving the $i$th bin from
collisions that goes into the $k$th bin, or the redistribution
function, which we assume to be scale independent. We assume that
fragments produced in catastrophic collisions have a range of masses
from the largest fragment, with $\frac{M_i}{2}$ labelled $i_{lr}$, to
the smallest body considered, labelled by $i_{max}$, which we assume
to be much smaller than $\frac{M_i}{2}$. Thus, the $k$th bin can only
gain mass from catastrophic collisions between objects with a mass
$2M_k$ or greater, labelled by $i_{mk}=k  - \frac{ln
  (2)}{\delta}$. Thus, the mass rate gained for solids in the $k$th
bin is calculated by summing over the contributions from the largest
mass bin, $i=1$, down to $i_{mk}$, which labels the bin of mass
$2M_i$. We assume that the scaling of the mass distribution of the fragments, $\alpha>1$ and that the logarithmic spacing between mass bins, $\delta<<1$. This leads to a redistribution function given by:
\begin{linenomath}
\begin{equation}
F_s(k-i)= (1-\delta)^{(k-i)(2-\alpha)}\delta (2-\alpha) 2^{(\alpha-2)}.
%pow((1-delta),(kminusi[i]*(2-alpha)))*delta*(2-alpha)* pow((1./2.),((alpha-2)))
%F_s(k-i)= (1-\delta)^{(k-i)(1-\alpha)}\delta^{(1-\alpha)} (1-\alpha) 2^{\alpha-1}
\end{equation}
\end{linenomath}
This is based on Eq.~20 of \cite{wyatt11}, where $\delta$ is now the spacing between mass bins and not radial bins, $\eta_{\rm max}=1/2$, such that $\delta = \delta'/3$ and $\alpha'=3 \alpha-2$, where $\delta'$ and $\alpha'$ are the parameters used in \cite{wyatt11}.

At every time-step, we use Eq.~\ref{eq:massconservation}, Eq.~\ref{eq:mloss}, Eq.~\ref{eq:mgain} to
track the mass gained and lost. We also track
$O_{k,i}$ which refers to the mass in the $k$th bin which originated
from the $i$th bin. At every time-step, each $j$th bin loses mass at
$m_{s,j}^c R_j^c \Delta$, a fraction $O_{j,i}$ of which originally
came from the $i$th bin. In order to keep track of the evolution of mass
that started the simulation in the $i$th bin, we calculate: 
\begin{linenomath}
\begin{equation}
O_{k,i} =\frac{ \left( O_{k,i}m_{s,k} - O_{k,i}m_{s,k}R_k^c \Delta + \Sigma_{j=1}^{j_{mk}}O_{j,i} F(k-j)m_{s,j} R_j^c \Delta\right)} {m_{s,k} - m_{s,k} R_k\Delta + \Sigma_{j=0}^{j_{mk}} F(k-j)m_{s,j} R_j^c \Delta},
\label{eq:oki}
\end{equation}
\end{linenomath}
and keep track of the mass originating in the $k$th bin, which remains in the $k$th bin, which is crucial for tracing the mass of material that has never been involved in collisions and thus, never changed bins: 
\begin{linenomath}
\begin{equation}
O_{k,k} = \frac{ O_{k,k}m_{s,k} - O_{k,k}m_{s,k}R_k^c \Delta }{m_{s,k} - m_{s,k} R_k\Delta + \Sigma_{j=0}^{j_{mk}} F(k-j)m_{s,j} R_j^c \Delta}
\label{eq:okk}
\end{equation}
\end{linenomath}
where the denominator is just the mass in the bin at the next time step. 
There should be no material in the bins with $i>i_{mk}$ and
the sum of $\Sigma_{i=1}^{i_{max}} O_{k,i}=1$ for conservation of
mass. As each bin loses mass ($m_{s,k}R_k^c$) every timestep, we assume that a fraction $O_{k,i}$ is lost from the material in $k$ originating from $i$.

\begin{figure}
\includegraphics[width=0.48\textwidth]{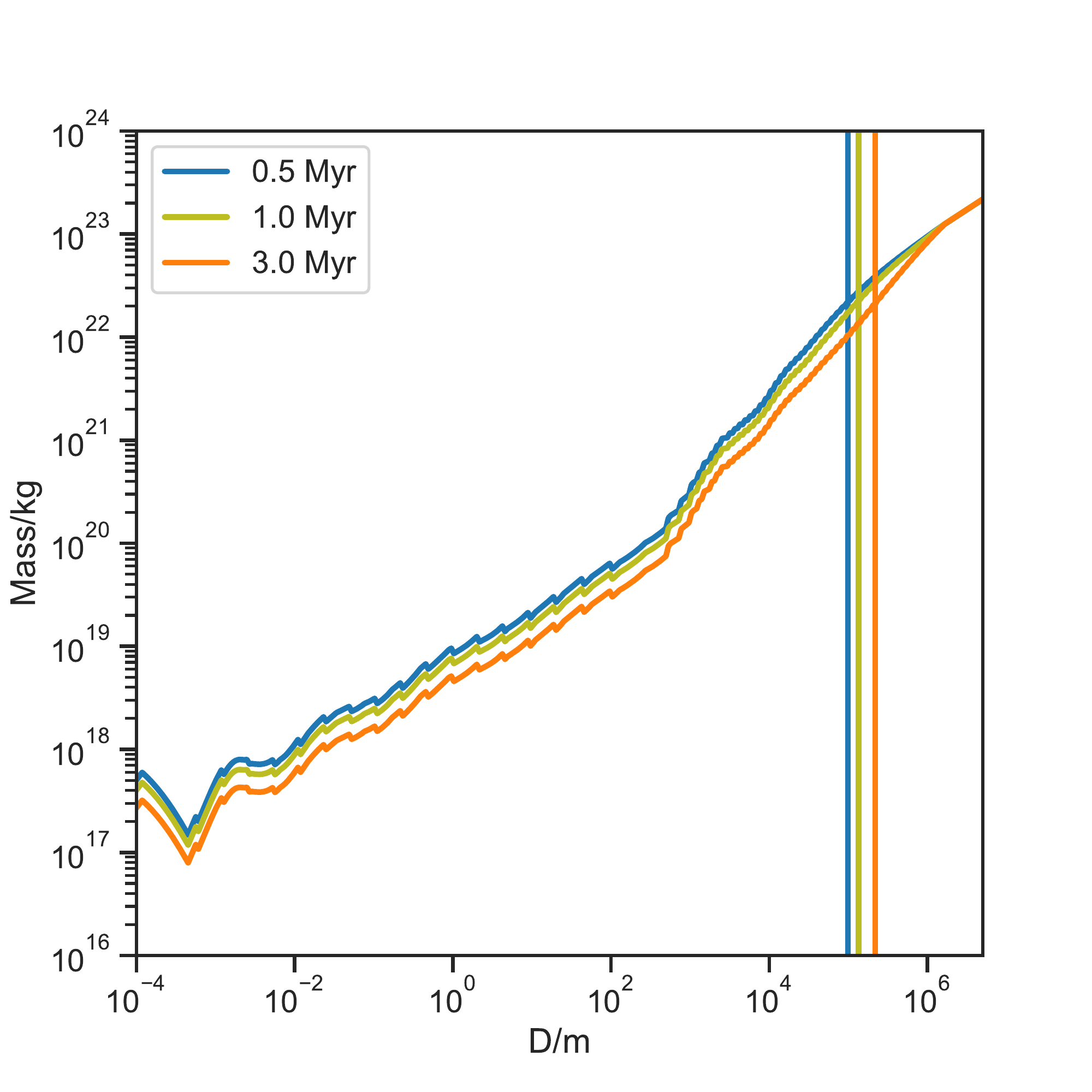} % ablet_1au_ecc_0.05_oct_2020
\includegraphics[width=0.48\textwidth]{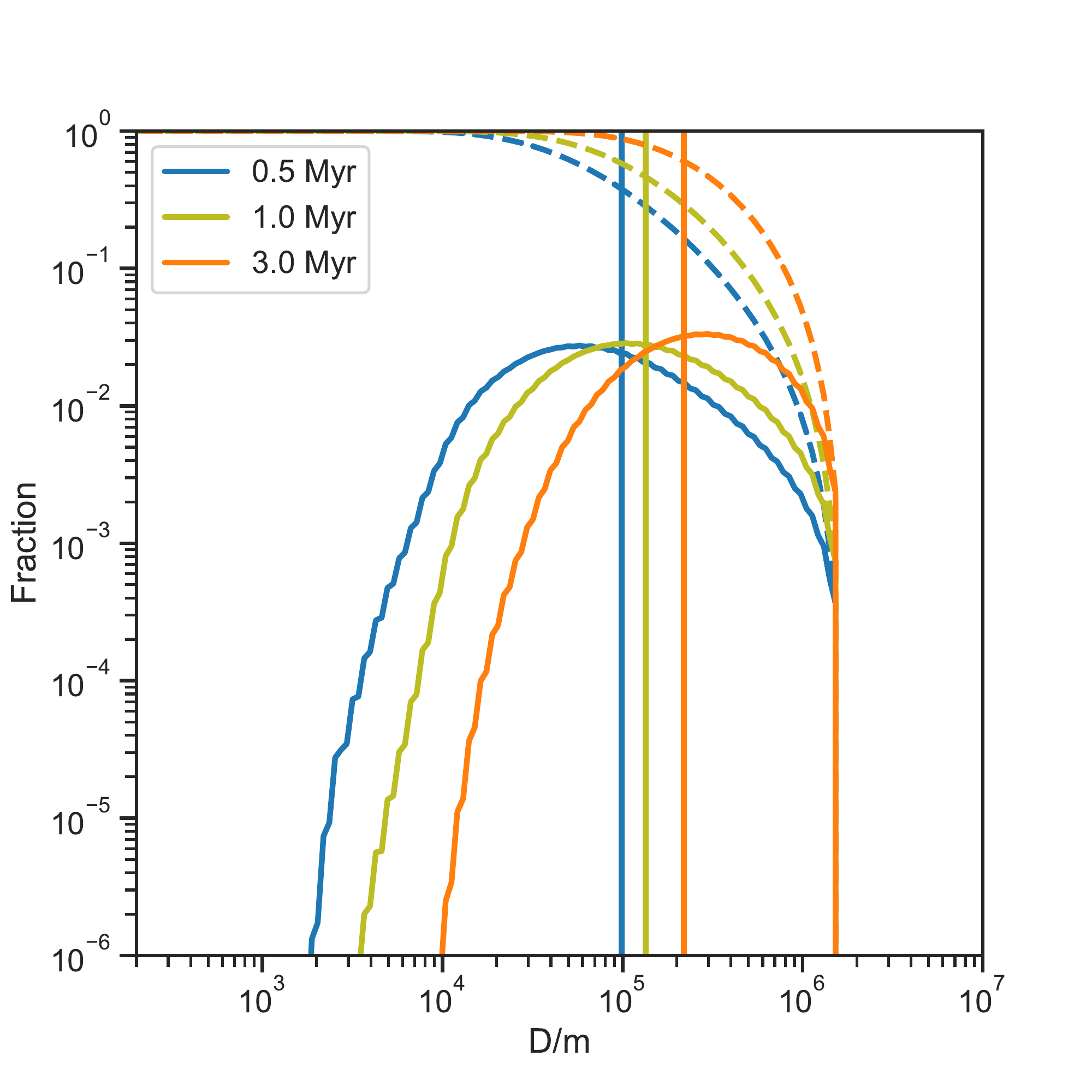}
\caption{{\bf The collisional evolution of the planetesimal belt.} Panel A: The size distribution of the planetesimal belt at 1 au, with an initial mass of $1 M_\oplus$ in bodies between 100$\mu$m and 5,000km, eccentricity $e=0.1$ and logarithmic spacing between mass bins $\delta=0.2$ and $\alpha=3.5$ at three epochs. Vertical lines indicate the estimated size for the bodies that have just entered collisional equilibrium ($D_{\rm eq}$, Eq.17). Panel B: The fraction of material in the 100 m-sized planetesimals that originates from all other bins at various epochs for a belt at 1 au, for the same simulation. The dashed lines indicate the cumulative sum and vertical lines label $D_{\rm eq}$ (Eq~\ref{eq:deq}). }
\label{fig:origin}
\end{figure}

\subsection{Simulations}
 Individual planetesimal belts are simulated by distributing mass between size bins, according to an initial size distribution, with $\alpha=3.5$ and logarithmic bins of
width $\delta=0.2$ (Eq.~\ref{eq:n_m}). The mass in each size bin is iterated forward in time according to Eq.~\ref{eq:massconservation}.  We fixed the belt width, $dr$ at 0.5, the particle's density at $3\times10^3$ kg m$^{-3}$ and consider belts with initially 100 $M_\oplus$ of material, at a radius of 1 au, with initial particle eccentricity of $0.1$. We consider particles with diameters between
$D_{\rm min}=100$ $\mu$m and 5,000 km (an arbitrary upper bound, which it
will be shown does not influence the results). The bin width and timestep are chosen to be sufficiently small that the
mass lost and gained by the smallest particles in one timestep are not
a significant fraction of the total mass in that bin, with $\delta_t= 10^6$s.  %A timestep of %$5\times 10^4$,
%$10^6$s is used. The bin width and timestep are chosen to be sufficiently small that the
%mass lost and gained by the smallest particles in one timestep are not
%a significant fraction of the total mass in that bin. 

The material in the belt is rapidly collisionally depleted. The smallest grains quickly reach collisional equilibrium, whilst the largest grains/planetesimals are unlikely to suffer collisions and retain their primordial size distribution. The left-hand panel of Supplementary Figure~1 shows the size distribution of an example planetesimal belt at 1 au. The apparent wave in the size distribution results from the grain cut-off at a single size for the smallest grains, as discussed in {\it e.g.} \cite{wyatt11, lohne}. Those bodies for whom the collisional lifetime is less than the age of the system are collisionally depleted ($t_c(D)<t$), whilst larger bodies are not collisionally evolved. For older systems, larger and larger bodies enter collisional equilibrium.

%Semi-analytic
%models for the collisional evolution of debris discs, calculate a
%value for $D_{\rm eq}$ by considering a simple form of the collision
%lifetime ($t_c$) of a body of diameter, $D$ \cite{wyatt03?, wyatt_review}: 
%\begin{equation}
%t_c=1.4 \times 10^{-9} \frac{(r/au)^{13/3} \frac{dr}{r} (D/km)
%(Q^*_D/ {\rm Jkg}^{-1})^{5/6}  }{e^{5/3}(M_*/M_\odot)^{4/3}  (M_{\rm tot}/M_\oplus)} 
%\, \, \, {\rm Myr}
%\end{equation}

If we consider a collision time of \cite{Wyatt07hot} (Eq.~7)
\begin{linenomath}
\begin{equation}
    t_{c (D)}= t_{\rm per} \frac{r dr}{\sigma_{\rm tot}} \frac{2 I }{f(e,I)} \frac{1}{f_{cc}}
\end{equation}
\end{linenomath}
where $f(e,I)$ is the ratio of the relative velocity of collisions to the Keplerian velocity ($v_{\rm rel}/v_k$), where $e$ and $I$ are the mean particle eccentricity and inclinations, $\sigma_{\rm tot}$ is the total cross-sectional area, $f_{cc}$ is the fraction of the total cross-sectional area in the belt which is seen by planetesimals of size $D$ as potentially causing a catastrophic collision. Following \cite{Wyatt07hot}, this can be written as: 
\begin{linenomath}
\begin{equation}
    f_{\rm cc} = \left(\frac{D_{\rm min}}{D}\right)^{3q-5} G(q, X_c),
\end{equation}
\end{linenomath}

where $G(q, X_c)$ is a function of both the size distribution (q) and the ratio of the smallest planetesimal ($D_{cc}$) that has enough energy to catastrophically destroy a planetesimal of size D, $X_c= D_{cc}/D$. This can be calculated in terms of the dispersal threshold, $Q_D^*$: 
\begin{linenomath}
\begin{equation}
X_c=\left(\frac{2Q_D^*}{v_{\rm rel}^2}\right)^{1/3}.
\end{equation}
\end{linenomath}
For a typical collisional cascade, $X_c<<1$, such that the function $G(q, X_c)$, for $q=11/6$,  can be approximated as $G(11/6, X_c) \sim 0.2X_c^{-2.5}$ for $X_c <<1$.  %% NB fix q=11/6 in future expressions! 
The total cross-sectional area can be related to the total disc mass ($M_{\rm tot}$)
\begin{linenomath}
\begin{equation}
    \frac{\sigma_{\rm tot}}{M_{\rm tot}} = \frac{3}{4 \rho} \frac{ D_{\rm min} ^{5-3q}}{D_{\rm max}^{6-3q}} \left( \frac{3q-6}{5-3q}\right)
\end{equation}
\end{linenomath}

%\begin{equation}
%    f_{\rm cc} = \left(\frac{D_{\rm min}}{D}\right)^{3q-5} G(q, X_c)
%\end{equation}

%\begin{equation}
%    \sigma_{\rm tot} f_{\rm cc}= \frac{3 M_{\rm tot} }{4 \rho D} G\frac{3q-6}{5-3q}
%\end{equation}
Thus, leading to an expression for the collisional lifetime of a particle of diameter, D: 
\begin{linenomath}
\begin{eqnarray}
    t_{c}& = & t_{\rm per} \frac{r dr}{\sigma_{\rm tot} }\frac{ 2 I }{f(e,I) } \frac{1}{f_{cc}}\\
    & = & t_{\rm per} \frac{r\, dr\, 4 \rho\, D}{3 M_{\rm tot}} \frac{ 2\, I}{ G(q, X_c)f(e, I)} \left (\frac{3q-5}{6-3q}\right).
    \label{eq:t_c}
\end{eqnarray}
\end{linenomath}

As time continues, larger and larger particles reach collisional equilibrium. The size particle that has just reached collisional equilbirium ($D_{\rm eq}$) can be approximated by the size particle for whom the collisional lifetime is equal to the current time $t = t_c (D_{\rm eq})$. In the regime where D is large ($D>800$m), the dispersal threshold, $Q_D^*$ (Eq.~\ref{eq:qdstar}) can be approximated as $Q_D^*\sim Q_b D^b$. Then $D_{eq}$ is given by 
\begin{equation} 
D_{\rm eq}=(t/K)^{1/(1+5b/6)},
  \label{eq:deq}
\end{equation}
where 
\begin{linenomath}
\begin{equation}
    K =  t_{\rm per} \frac{0.2 r dr 4 \rho}{3 M_{\rm tot}} \frac{2I}{f(e,I)} \left(\frac{v_{\rm rel^2}}{2 Q_b}\right) ^{5/6}. \nonumber
\end{equation}
\end{linenomath}
% $t = K D ^(1+5b/6)$

We note here that this size is an approximation and that the absence of small grains leads to a size distribution that deviates from a perfect power law  (see Supplementary Figure~1.)

\subsection{The collisional cascade is fed by the largest bodies}

The bodies that have just reached collisional equilibrium ($D_{\rm eq}$) dominate the mass evolution of the belt \citep{wyatt11}. Here we trace the origin of the material arriving in each size bin, using Eq.~\ref{eq:oki}, ~\ref{eq:okk}, with the aim of investigating the extent to which the bodies that have just reached collisional equilibrium dominate the mass budget in small bodies. The smallest bodies are continuously lost from the collisional cascade, and thus, new material must replenish bodies of all sizes.

The right-hand panel of Supplementary Figure~1 shows the fraction of the mass in the
diameter bin centred on $D_k=100$m
that originated from larger diameters. The $D_k=100$m was chosen to represent any particles that are fully in collisional equilibrium and constantly being resupplied by collisions between larger bodies. The mass budget is indeed dominated by bodies of around $D_{\rm eq}$, as shown by the vertical lines. $D_{\rm eq}$ as calculated by Eq.~\ref{eq:deq} is an approximation, not taking into account the wavy nature of the size distribution and does not perfectly calculate the true maximum size in collisional equilibrium (see Supplementary Figure~1), nor align perfectly with the maximum here, but the approximation is good to within a factor of a few.

The right-hand panel of Fig.~4 shows the fraction of material in the smaller size grains that originates from grains larger than a certain size, $1,400$km, as a function of time, plotted in units of the collisional lifetime of these largest bodies ($t_c(D=1,400km)$. As the bodies enter collisional equilibrium, they dominate the mass in smaller size bins, but the mass in small bodies ($D_{\rm in}$ from $D>1,400$km tends to one only on timescales longer than the collision timescale. The fraction of material from $D>1,400$km in 30km planetesimals reaches a  percent after 0.1$t_c(D_*)$. This would apply to the fraction of planetesimals, $D>D_*$ in the bin labelled by $D_*/50$, where in this case $D_*=1,400$km.

The form of right-hand panels of Supplementary Figure~1 and Fig.~4 remain similar for different diameters and we assert that within the validity of the approximation for $t_c$ and accounting for small differences due to the wavy nature of the size distribution, the form of these figures is independent of the sizes $D_*$ and $D_{\rm in}$. Any differences result from the wavy nature of the size distribution and the approximations used in $t_c(D)$, whose validity change with diameter. The self-similar nature of the collisional cascade saves us from needing to run the collisional model on sufficiently long timescales that bodies of $>1,400$ km enter collisional equilibrium.

\subsection{Frequency of Pluto-fed polluted white dwarfs}

Although planetesimal belts sufficiently massive and sufficiently close-in that even the largest ($D>1,400$ km) planetary bodies are collisionally evolving are rare, the aim of the following section is to assess whether they are sufficiently common to explain core(mantle)-rich compositions in some pollutants of white dwarfs. In this scenario, no \al would be required to form an iron core.

Assuming that all planetesimal belts contain bodies larger than 1,400 km, the properties of those planetesimal belts in which large ($D>1,400$ km) bodies would be collisionally evolving can be estimated by considering a typical lifetime for the planetary system. Many white dwarfs evolved from main-sequence A stars, where typical main-sequence lifetimes are on the order of hundreds of Myrs. Belt radii expand by a factor of 2-3 during the white dwarf phase, following mass loss, so the majority of the collisional evolution occurs during the main-sequence phase \citep{bonsor10}. For solar-type stars, main-sequence lifetimes can be as long as tens of Gyrs, but the age of the Universe stipulates that very few white dwarfs had main-sequence lifetimes this long. Thus, we consider a conservative estimate on the timescale for which collisional evolution occurred of 5 Gyr. Using a typical distribution of planetesimal belts, fitted to observations of debris discs around main-sequence A stars \citep{wyatt07}, with the distribution of initial belt radii is $n(r)dr\propto r^\gamma dr$, with $\gamma=-0.8$, between 3 and 200 au, the distribution of initial belt masses forms a log normal distribution of width 1.13 dex, centred on $10 M_\oplus$ of width $M\oplus$, we find that a few tenths of a percent of belts have a collisional lifetime for particles of size 1,400 km less than 5 Gyr. About a percent of systems have 10\% of the collisional lifetime of $D=1,400$ km less than 5 Gyr. Planetary systems in which such large bodies are catastrophically colliding are rare. Thus, planetary systems where 10-100 km planetesimals are likely to be the collision fragments of larger core--mantle differentiated Plutos are rare. Additionally, only a sub-set of collision fragments will have core or mantle compositions sufficiently extreme to be detected. If this fraction is on the order of 10\% (see \emph{e.g.} Fig.~3 of \cite{Bonsor2020}), we anticipate that core- or mantle-rich compositions would show up in $<<0.05\%$ of white dwarfs without the need for \al. Thus, only a tiny fraction of white dwarf pollutants are likely to originate from large bodies, as this fraction is significantly lower than the fraction of white dwarf pollutants that appear to be core(mantle)-rich of at least 4\% (see ~\ref{sec:sampleone}, \ref{sec:sampletwo}). 

Additionally, the existence of large bodies in planetesimal belts has been placed in question \citep{Krivov_Wyatt2021}, and if such large bodies do exist, it is not clear that they would have the same size distribution as the rest of the belt. However, it is plausible that in some planetary systems dynamical instabilities lead to high velocity collisions or excite collisions in planetesimals belts outside of the normal steady-state collisional evolution considered here.

\bmhead{Data Availability}
The data used to create all figures is available in the Supplementary Information, most notably the white dwarf data (Sample One) is detailed in Extended Data Tables 1, 2 and 3, whilst Sample Two is found in \citep{Harrison2021}.

\bmhead{Code Availability}
The code used to create all figures and the collisional evolution code is available at \url{ https://github.com/abonsor/collcascade }, which links to models available at \url {https://github.com/timlichtenberg/2stage_scripts_data} for Figure 2.

\bmhead{Acknowledgments}
A.B. acknowledges support from a Royal Society Dorothy Hodgkin Research Fellowship, DH150130 and a Royal Society University Research Fellowship, URF$\backslash$R1$\backslash$211421.
T.L. was supported by a grant from the Simons Foundation (SCOL award No. 611576). J.D. acknowledges funding from the European Research Council (ERC) under the European Unions Horizon 2020 research and innovation programme under grant agreement No.~714769. A.M.B. acknowledges support from a Royal Society funded PhD studentship, RGFEA180174. We acknowledge fruitful discussions with Marc Brouwers, Laura Rogers, Elliot Lynch, Alfred Curry, Til Birnstiel,  Mark Wyatt, and Richard J. Parker.

\bmhead{Author Contributions Statement}

The idea for the study came from discussions between A.B., J.D. and T.L. The analysis of the white dwarf data was performed by A.M.B., whilst T.L. supplied the thermal evolution models used for Fig 2. The manuscript was written in collaboration between all authors.  

\bmhead{Competing Interests Statement}

The authors declare no competing interests.

%%===========================================================================================%%
%% If you are submitting to one of the Nature Portfolio journals, using the eJP submission   %%
%% system, please include the references within the manuscript file itself. You may do this  %%
%% by copying the reference list from your .bbl file, paste it into the main manuscript .tex %%
%% file, and delete the associated \verb+\bibliography+ commands.                            %%
%%===========================================================================================%%

%\bibliographystyle{sn-standardnature}
%\bibliography{ref,ref2}% common bib file
%% if required, the content of .bbl file can be included here once bbl is generated

%% Default %%
%%\input sn-sample-bib.tex%

%%%%%%%%%%%%%%%%%%%%%%%%%%%%%%%%%%%%%%%%%%%%%%%%%%%%%%%%%%%%%%%%%%%%%%%%%%%%%%%%%%%%%%%%%%
%%%%%%%%%%%  SUPPLEMENTARY MATERIALS
%%%%%%%%%%%%%%%%%%%%%%%%%%%%%%%%%%%%%%%%%%%%%%%%%%%%%%%%%%%%%%%%%%%%%%%%%%%%%%%%%%%%%%%%%%

\end{document}